\documentclass[twocolumn, english]{article}

\usepackage[sorting = none, backend = biber, style=numeric-comp]{biblatex}
\usepackage{graphicx}
\usepackage{hyperref}
\usepackage{authblk}

\usepackage{geometry}
\begin{document}

\title{Wavefront shaping in multimode fibers by transmission matrix engineering}

\author[1]{Shachar Resisi}
\author[1]{Yehonatan Viernik}
\author[2]{Sebastien Popoff}
\author[1,*]{Yaron Bromberg}

\affil[1]{Racah Institute of Physics, The Hebrew University of Jerusalem, Israel}
\affil[2]{Institut Langevin, CNRS UMR 7587, ESPCI Paris, PSL Research University, 1 rue Jussieu, 75005 Paris, France}
\affil[*]{Corresponding author: yaron.bromberg@mail.huji.ac.il}

\date{}

\twocolumn[\begin{@twocolumnfalse}
\maketitle


\begin{abstract}
One of the greatest challenges in utilizing multimode optical fibers is mode-mixing and inter-modal interference, which scramble the information delivered by the fiber. 
A common approach for canceling these effects is to tailor the optical field at the input of the fiber to obtain a desired field at its output.
In this work, we present a new approach which relies on modulating the transmission matrix of the fiber rather than the incident light. 
We apply computer-controlled mechanical perturbations to the fiber to obtain a desired intensity pattern at its output.
Using an all-fiber apparatus, we demonstrate focusing light at the distal end of the fiber and conversion between fiber modes. 
Since in this approach the number of degrees of control can be larger than the number of fiber modes, it allows simultaneous control over multiple inputs and multiple wavelengths. 
\newline
\end{abstract}

\end{@twocolumnfalse}]


\section{Introduction}

In recent years, multimode optical fibers (MMFs) are at the focus of numerous studies aiming at enhancing the capacity of optical communications and endoscopic imaging systems \cite{Richardson2013, Ploschner2015}. 
Ideally, one would like to utilize the transverse modes of the fiber to deliver information via multiple channels, simultaneously. However, inter-modal interference and coupling between the guided modes of the fiber result in scrambling between channels. 
One of the most promising approaches for unscrambling the transmitted information is by shaping the optical wavefront at the proximal end of the fiber in order to get a desired output at the distal end. 
Demonstrations include compensation of modal dispersion \cite{Shen2005, Alon2014,Wen2016}, focusing at the distal end \cite{DiLeonardo2011,Papadopoulos2012, Caravaca-Aguirre2013,Papadopoulos2013,BoonzajerFlaes2018}, and delivering images \cite{Cizmar2012, Bianchi2012, Choi2012} or an orthogonal set of modes \cite{Carpenter:14, Carpenter2015} through the fiber. 

\begin{figure}[!tbh]
\begin{centering}
\includegraphics[width=\columnwidth]{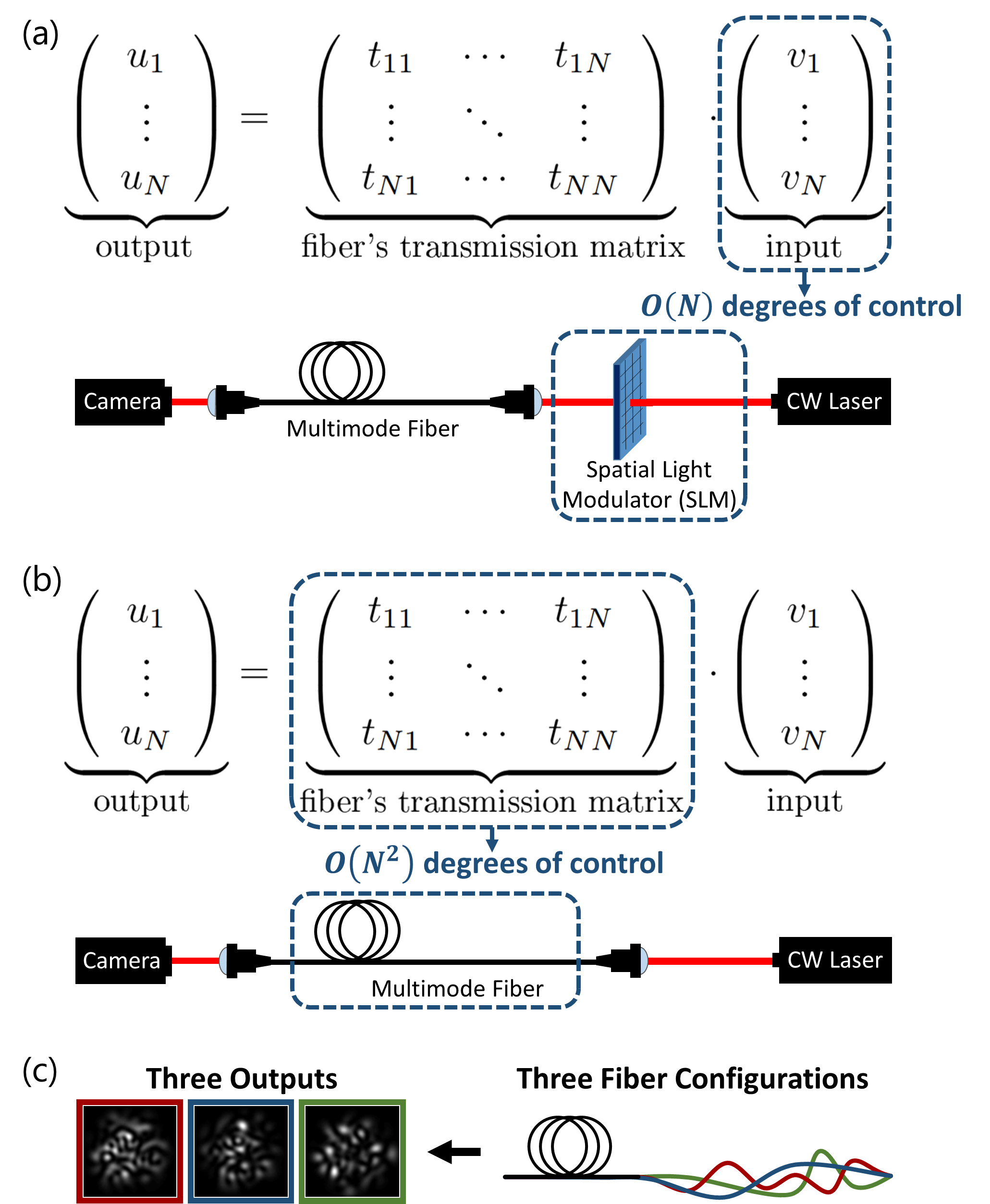}
\par\end{centering}
\caption{\textbf{Shaping the transmission matrix of multimode optical fibers.}
(a) The conventional method for wavefront shaping in complex media, performed e.g. by using an SLM and free space optics to tailor the incoming wavefront at the proximal end of the multimode fiber. (b) Proposed method for light modulation, in which the transmission matrix of the medium is altered, e.g. by performing perturbations on the fiber itself. (c) Illustration of the sensitivity of the output pattern on the fiber geometry. Three different configurations of the fiber (depicted by red, green and blue curves), correspond to three different speckle patterns at the output of the fiber. Since the input field coupled into the fiber is fixed, the different output patterns correspond to different transmission matrices of the fiber.}
\label{fig:matrix}
\end{figure}

Typically in wavefront shaping, the incident wavefront is controlled using spatial light modulators (SLMs), digital micromirror devices (DMDs) or nonlinear crystals.
In all cases, the shaped wavefront sets the superposition of guided modes that is coupled into the fiber. For a fixed transmission matrix (TM) of the fiber, this superposition determines the field at the output of the fiber, as depicted in Fig. \ref{fig:matrix}(a)). Hence, in a fiber that supports $N$ guided modes, wavefront shaping provides at most $N$ complex degrees of control. 
However, many applications require the number of degrees of control to be larger than the number of modes. 
For example, one of the key ingredients for spatial division multiplexing is mode converters, which require simultaneous control over the output field of multiple incident wavefronts.
To this end, complex multimode transformations were previously demonstrated by applying phase modulations at multiple planes \cite{Morizur:10, Labroille:14, Fontaine2019, Fickler2019}. However, this requires free-space propagation between the modulators, thus limiting the stability of the system and increasing its footprint. 

In this work we propose and demonstrate a new method for controlling light at the output of MMF, which does not rely on shaping the incident light and that can be implemented in an all-fiber configuration. 
Inspired by the ongoing efforts to generate on-chip mode converters by manipulating modal interference in multimode interferometers \cite{Piggott2015, Bruck2016, Harris:18}, we directly control the light propagation inside the fiber to manipulate its TM, allowing us to generate a desired field at its output (Fig. \ref{fig:matrix}(b)). Since the TM is determined by $O\left(N^{2}\right)$ complex parameters, TM-shaping provides access to much more degrees of control than shaping the incident wavefront. 

To control the fiber's TM, we apply computer controlled bends at multiple positions along the fiber. 
Since the stress induced by the bends changes the boundary conditions of the system, it modifies the TM such that different bends yield different speckle patterns at the distal end (Fig. \ref{fig:matrix}(c)). 
We can therefore obtain a desired field at the output of the fiber by imposing a set of controlled bends, without modifying the incident wavefront. 
Since in this approach the input field is fixed, it does not require an SLM or any other free-space component. Such an all-fiber configuration is especially attractive for MMF-based applications that require high throughput and an efficient control over the field at the output of the fiber. As a proof-of-concept demonstration of TM-shaping, we demonstrate focusing at the distal end of the fiber, and conversion between the fiber modes. 

\section{Experimental Techniques}

\subsection{Principle}

Our method relies on applying controlled weak local bends along the fiber. To this end, we use an array of computer-controlled piezoelectric actuators to locally apply pressure on the fiber at multiple positions \cite{Golubchik2015, regelman2016method}. 
The TM of the fiber depends of the curvatures of the bends, which are determined by the travel of each actuator. 
To obtain a target pattern at the distal end, we compare the intensity pattern recorded at the output of the fiber with a desired target pattern. Using an iterative algorithm, we search for the optimal configuration of the actuators, i.e. the optimal travel of each actuator, that maximizes the overlap of the output and target patterns. 
 
\subsection{Experimental Setup}
\begin{figure}[!tbh]
\begin{centering}
\includegraphics[width=\columnwidth]{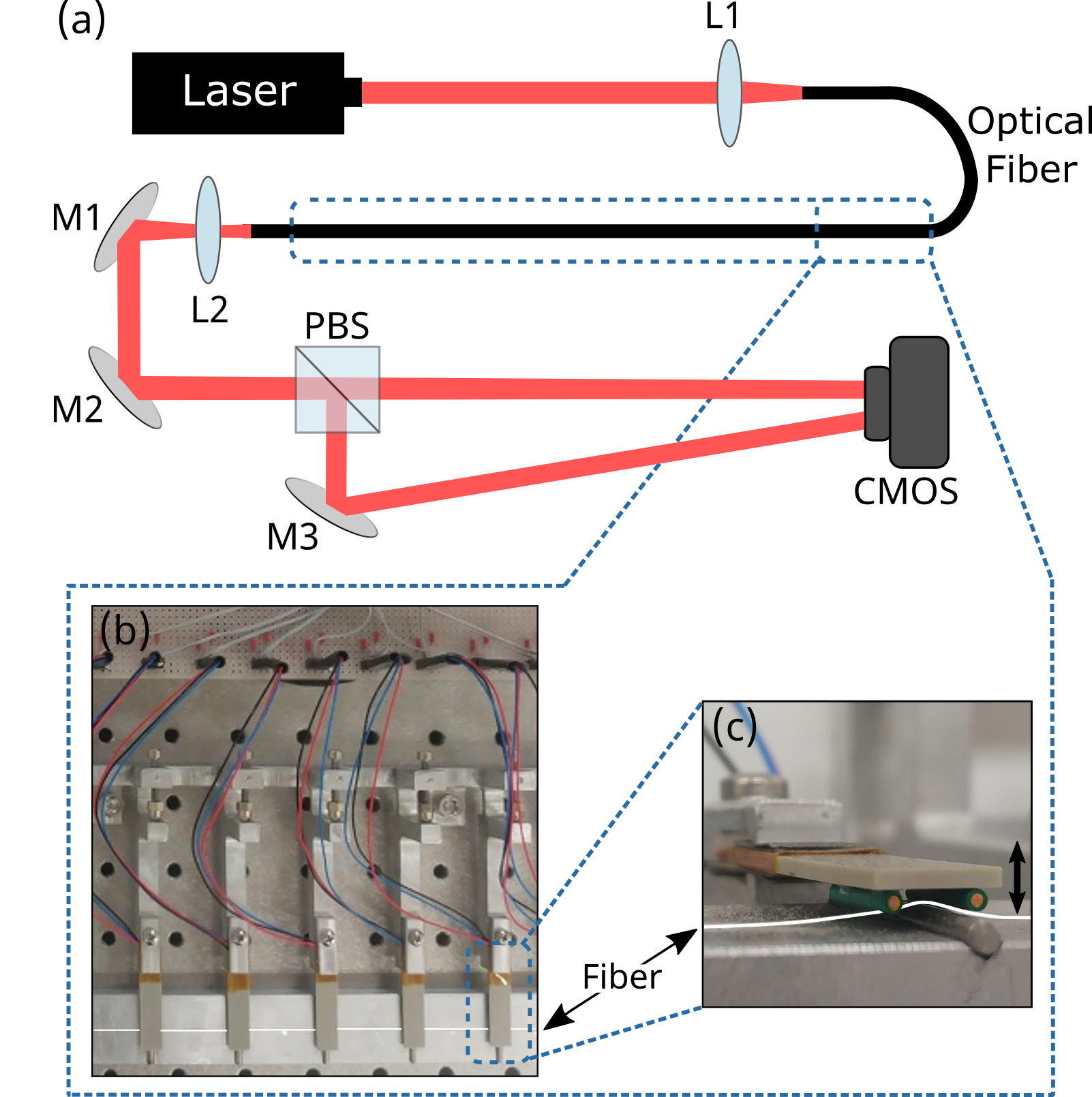}
\par\end{centering}
\caption{\textbf{Experimental setup for controlling the transmission matrix of optical fibers.} 
(a) The laser beam is coupled into the optical fiber, which is fixed to a metal bar. 37 actuators are placed above the fiber, applying local vertical bends. The light that is emitted from the distal facet of the fiber travels through a polarizing beamsplitter, and both horizontal and vertical polarizations are recorded by a CMOS camera. (b) Top view of five actuators, bending the fiber from above. (c) The fiber is pressed by two pins that are attached to each actuator, and one pin which is placed below it, creating a three-point contact. A computer-controlled voltage that is applied on each actuator sets its travel and defines the curvature of the local deformation it poses on the fiber. L, lens; M, mirror; PBS, polarizing beamsplitter; CMOS, camera.}
\label{fig:setup}
\end{figure}

The experimental setup is depicted in Fig. \ref{fig:setup}. A HeNe laser (wavelength of $\lambda=632.8 \hspace{2pt} nm$) is coupled to an optical fiber, overfilling its core. We placed 37 piezoelectric actuators along the fiber. By applying a set of computer-controlled voltages to each actuator, we controlled the vertical displacement of the actuators.  
Each actuator bends the fiber by a three-point contact, creating a bell-shaped local deformation of the fiber, with a curvature that depends on the vertical travel of the actuator (see Figs. \ref{fig:setup}(b,c)). 
For the maximal curvature we applied ($R\approx10 \hspace{2pt} mm$), we measured an attenuation of few percent per actuator due to bending loss. The spacing between nearby actuators was set to be at least $3 \hspace{2pt} cm$, which is larger than $\frac{d^2}{\lambda}$ for $d$ the core's diameter, such that the interference pattern inside the fiber between two adjacent actuators is uncorrelated. At the distal end, a CMOS camera records the intensity distribution of both the horizontally and vertically polarized light.  

We used two types of multimode fibers: a fiber supporting few modes for demonstrating mode conversion, and a fiber supporting numerous modes for demonstrating focusing. For the focusing experiment, we used a 2 meter-long graded-index (GRIN) multimode optical fiber with numerical aperture (NA) of 0.275 and core diameter of $d_{MMF}=62.5  \hspace{2pt} \mu m$ (InfiCor OM1, Corning). The fiber supports approximately 900 transverse modes per polarization  at $\lambda=632.8 \hspace{2pt} nm$ ($V\approx85$), yet we used weak focusing at the fiber's input facet to excite only $\approx 280$ modes.
For the experiments with the few mode fiber (FMF), we used a 5 meter-long step-index (SI) fiber, with an NA of 0.1 and core diameter of $d_{FMF}=10  \hspace{2pt} \mu m$ (FG010LDA, Thorlabs). In principle, at our wavelength the fiber supports 6 modes per polarization, ($V \approx5$).

\subsection{Optimization Process}

The curvature of the bends, set by the travel of each actuator, modifies how light propagates through the fiber and thus determines the speckle pattern that is received at the distal end. 
We can therefore define an optimization problem of finding the voltages that should be applied to the actuators, to receive a given target pattern at the output of the fiber. 
The distance between the target and each measured pattern is quantified by a cost function, which the algorithm iteratively attempts to minimize. 

For $M$ actuators, the solution space is an $M$-dimensional sub-space, defined by the voltages range and the algorithm's step intervals, and can be searched using an optimization algorithm. While the optical system is linear in the optical field, the response of the actuators, i.e. the modulation they pose on the complex light field, is not linear in the voltages.

Moreover, since a change in the curvature of an actuator at one point along the fiber affects the interference pattern at all of the following actuators positions, the actuators cannot be regarded as independent degrees of control. 
Similar nonlinear dependence between degrees of control is obtained, for example, for wave control in chaotic microwave cavities \cite{Geoffroy2018}. 
Out of the wide range of iterative optimization algorithms that can efficiently find a solution to such nonlinear optimization problems, we chose to use Particle Swarm Optimization (PSO) \cite{PSO}, as on average it achieved the best results out of the algorithms we tested (See the Supplementary Material for more details regarding the use of PSO).

\section{Results}
\subsection{Focusing at the Distal End of the Fiber}

\begin{figure}[!tbh]
\begin{centering}
\includegraphics[width=\columnwidth]{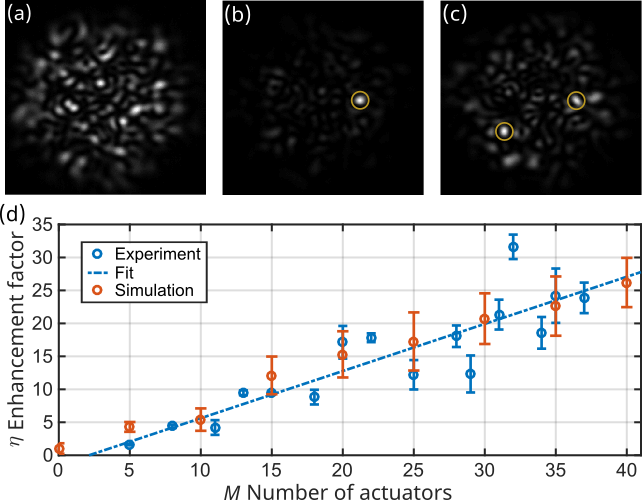}
\par\end{centering}
\caption{\textbf{Focusing at the output of a multimode fiber}. (a) Image of the speckle pattern at the output of the fiber, before the optimization process. 
(b, c) The output intensity pattern, after optimizing the travel of the 37 actuators to focus the light to a single target (b), and two foci simultaneously (c). (d) The average enhancement as a function of the number of active actuators. Each data point (blue circles) was obtained by averaging the enhancement over several experiments, where the error bars indicate the standard error.
A linear fit yields a slope of $\alpha\approx0.71$, which is close to the theoretical slope for phase-only modulation. The fit intersects the $\hat{y}$ axis at $M_{0}\approx1.5$, matching our observation that about $4-5$ actuators are required to overcome the inherent noise of the system. Numerical simulations for a GRIN fiber with $NA=0.275$ and $a=17.1 \hspace{2pt} \mu m$ (red circles), exhibit a linear scaling with a slope of $\alpha\approx 0.64$. The slope increases with the number of guided modes assumed in the simulation. Here we chose the number of modes ($M=280$) according to the number of excited modes in the experiment.  
}
\label{fig:MMF}
\end{figure}

To illustrate the concept of shaping the intensity patterns at the output of the fiber by controlling its TM, we first demonstrate focusing the light to a sharp spot at the distal end of the fiber. We excite a subset of the fiber modes by weakly focusing the input light on the proximal end of the fiber. Due to inter-modal interference and mode mixing, at the output of the fiber the modes interfere in a random manner, exhibiting a fully developed speckle pattern (Fig. \ref{fig:MMF}(a)). Based on the number of speckle grains in the output pattern, we estimate that we excite the first 280 guided fiber modes.

To focus the light to some region of interest (ROI) in the recorded image, we run the optimization algorithm to enhance the total intensity at the target area. We define the enhancement factor $\eta$ by the total intensity in the ROI after the optimization, divided by the ensemble average of the total intensity in the ROI before the optimization. The ensemble average is computed by averaging the output intensity over random configurations of the actuators, and applying an additional azimuthal integration to improve the averaging.

We start by choosing an arbitrary spot in the output speckle pattern of one of the polarizations. 
We define a small ROI surrounding the chosen position, in an area that is roughly the area of a single speckle grain, and run the optimization scheme to maximize the total intensity of that area.
Fig. \ref{fig:MMF} depicts the output speckle pattern of the horizontal polarization before (Fig. \ref{fig:MMF}(a)) and after (Fig. \ref{fig:MMF}(b) the optimization, using all 37 actuators. The enhanced speckle grain is clearly visible and has a much higher intensity than its surroundings, corresponding to an enhancement factor of $\eta=25$. 

We repeat the focusing experiment described above with a varying number of actuators $M$. When a subset of actuators is used, the remaining are left idle throughout the optimization. Fig. \ref{fig:MMF}(d) summarizes the results of this set of experiments, showing the obtained enhancement factor $\eta$ grows linearly with the number of active actuators $M$. 
It is instructive to compare this linear scaling with the well-known results for focusing light through random media using SLMs or DMDs.
Vellekoop and Mosk have shown that when the number of degrees of control (i.e. independent SLM or DMD pixels) is small compared to the effective number of transverse modes of the sample, the enhancement scales linearly with the number of degrees of control. The slope of the linear scaling $\alpha$ depends on the speckle statistics and on the modulation mode \cite{Vellekoop2007,Vellekoop2008,Vellekoop:15}. For Rayleigh speckle statistics, as in our system (see Supplementary Material), the slopes predicted by theory are  $\alpha=1$ for perfect amplitude and phase modulation, $\alpha=\frac{\pi}{4}\approx0.78$ for phase-only modulation \cite{Vellekoop:15}. Experimentally measured slopes, however, are typically smaller, mainly due to technical limitations such as finite persistence time of the system, unequal contribution of the degrees of control, and statistical dependence between them. 
Interestingly, we measure a slope of $\alpha\approx 0.71$, which is close to the theoretical value for phase-only modulation for Rayleigh speckles, and higher than typical experimentally measured slopes (e.g. $\alpha\approx 0.57$ in \cite{VellekoopPhdThesis}). Naively, one could expect a lower slope in our system, since as discussed above, in our configuration the degrees of control are not independent. The large slope values we obtain may indicate that the bends change not only the relative phases between the guided modes (corresponding to phase modulation), but also their relative amplitudes (corresponding to amplitude modulation), via mode-mixing and polarization rotation.

To further study the linear scaling, we performed a set of numerical simulations. We used a simplified scalar model for the light propagating in a GRIN fiber, in which the fiber is composed of multiple sections, where each section is made of a curved and a straight segment. The curved segments simulate the bend induced by an actuator, and the straight segments simulate the propagation between actuators (see Supplementary Material for more details). As in the experiment, we use the PSO algorithm to focus the light at the distal end of the fiber. The numerical results exhibit a clear linear scaling, with slopes in the range of 0.57-0.64 (see Fig. S3 in Supplementary Material). Simulations for fibers supporting $N=280$ modes, roughly the number of modes we excite in our experiment, exhibit a slope of $\alpha\approx0.64$, slightly lower than the the experimentally measured slope. 

As in experiments with SLMs, focusing is not limited to a single spot. To illustrate this, we used the optimization algorithm to simultaneously maximize the intensity at two target areas. Fig. \ref{fig:MMF}(c) shows a typical result, exhibiting an enhancement which is half of the enhancement obtained when focusing to a single spot, as expected by theory \cite{Vellekoop2008}. In principle, it is possible to focus the light to an arbitrary number of spots, yet in practice we are limited by the number of available actuators. 

\subsection{Mode Conversion in a Few Mode Fiber}

\begin{figure}[!tbh]
\begin{centering}
\includegraphics[width=0.8\columnwidth]{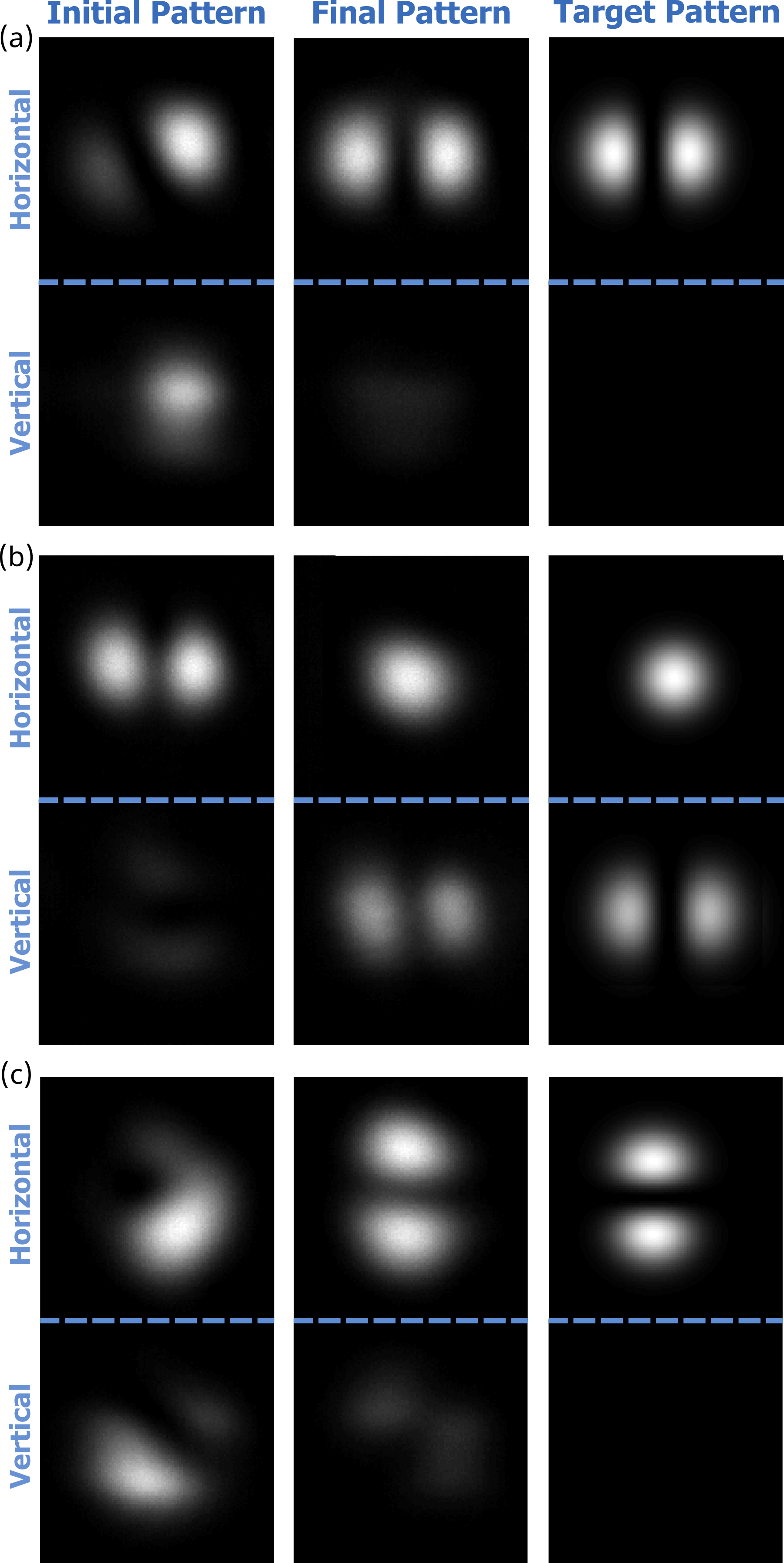}
\par\end{centering}
\caption{\textbf{Conversion between transverse fiber modes.} 
Intensity patterns recorded at the output of the fiber, before (left column) and after (middle column) the optimization, exhibiting conversion between the $LP$ fiber modes at orthogonal polarizations. The PSO algorithm iteratively minimizes the $\ell_{1}$ distance between the measured pattern and the target mask (right column). (a) and (b) are obtained using 33 actuators (with Pearson correlation of $0.94$ and $0.92$ respectively), (c) is obtained with 12 actuators (with correlation of $0.90$).}
\label{fig:FMF}
\end{figure}

In the previous section, we demonstrate the possibility to use our system as an all-fiber SLM, i.e. to shape an output complex field by modifying the relative complex weight of the propagating modes. In the following, we show that we can go further by studying the feasibility of TM-shaping to tailor the output patterns in the few-mode regime, where the number of fiber modes is comparable with the number of actuators. Specifically, we are interested in converting an arbitrary superposition of guided modes to one of the linearly-polarized (LP) modes supported by the fiber. To this end, we utilize the PSO optimization algorithm to find the configuration of actuators that maximizes the overlap between the output intensity pattern and the desired LP mode.
The target LP modes of the step-index fiber were computed numerically for the parameters of our fiber, and scaled to match the transverse dimensions of the fiber image. 
Fig. \ref{fig:FMF} presents a few examples of conversions between LP modes using 33 and 12 actuators. A mixture of $LP_{01}$ and $LP_{11}$ at two different polarizations can be converted to $LP_{11}$ in one polarization (Fig. \ref{fig:FMF}(a)). Alternatively, a horizontally polarized $LP_{11}$ mode can be converted to a superposition of horizontally $LP_{01}$ and a vertically polarized $LP_{11}$ (Fig. \ref{fig:FMF}(b)). The Pearson correlation between the target and final patterns in these examples is $0.93$. Similar results are obtained when we run the optimization with fewer active actuators, with a negligible reduction in the correlation between the target and final pattern. For example, with 12 actuators we observe correlations of $0.90$ for the conversion presented in Fig. \ref{fig:FMF}(c). Optimization with less than 12 actuator shows poorer performance, as the number of actuators becomes comparable with the number of guided modes.

\section{Discussion}
Controlling the transmission matrix of a multimode fiber, rather than the wavefront that is coupled to it, opens the door for an unprecedented control over the light at the output of the fiber. Since the number of degrees of control, the number of actuators in our implementation, is not limited by the number of fiber modes $N$, it can allow simultaneous control for orthogonal inputs and/or spectral components. 
In fact, if $O(N^2)$ degrees of control are available, one can expect generating arbitrary  $N \times N$ transformations between the input and output modes. Over the past two decades there is an ever-growing interest in realizing reconfigurable multimode transformations, for a wide range of applications, such as quantum photonic circuits \cite{Reck:1994, Carolan2015, Taballione2018, Harris:18, Gigan2019} optical communications \cite{Miller2015, Fontaine2019}, and nanophotonic processors \cite{Piggott2015, Annoni2017}. 
These realizations require strong mixing of the input modes, as the output modes are arbitrary superpositions of the input modes. The mixing can be achieved, for example, by diffraction in free-space propagation between carefully designed phase plates \cite{Morizur:10, Labroille:14, Fontaine2019, Fickler2019}, a mesh of Mach-Zehnder interferometers with integrated modulators \cite{Harris:18}, engineered scattering elements in multimode interferometers \cite{Piggott2015, Bruck2016}, or scattering by complex media \cite{Geoffroy2018, Maxime:19}. In our implementation, we rely on the natural mode mixing and inter-modal interference in multimode fibers, allowing implementation using standard commercially available fibers. 

The main limitation our current proof-of-concept suffers from is the achievable modulation rates.
The piezo-based implementation limits the achievable modulation rates. The response time of the system to abrupt changes of the piezos is approximately $30 \hspace{2pt} ms$ (see Supplementary Material), allowing in principle for modulation rates as high as 30 Hz. In practice, our system works at slower rates ($\approx$5 Hz), mainly due the latency of the piezoelectric actuators and the camera. The total optimization time for the focusing experiments is $50$ minutes, and $12-15$ minutes for the mode conversion experiments.
Faster electronics and development of a stiffer and more efficient bending mechanism will allow higher modulation rates, limited by the resonance frequency of the piezo benders ($\approx$ 300-500 Hz). To achieve even faster rates, a different technology should be used for applying perturbations to the fibers, e.g. utilizing all-fiber acousto-optical modulators \cite{acousto-optic-book} or the 'smart fibers' technology with integrated modulators \cite{Fink12}. Optical fibers with built-in modulators can also be utilized for a scalable implementation of our method.

\section{Conclusions and Outlook}

In this work we proposed a novel technique for controlling light in multimode optical fibres, by modulating its TM using controlled perturbations. We presented proof-of-principle demonstrations of focusing light at the distal end of the fiber, and conversion between guided modes, without utilizing any free-space components. 
Since our approach to modulate the TM of the fiber is general and not limited to mechanical perturbations, it could be directly transferred to other types of actuators, e.g. in-fiber electro-optical or acousto-optical modulators, to achieve all-fiber, loss-less, fast, and scalable implementations. 
The all-fiber configuration and the possibility to control more degrees of freedom than the number of guided modes, makes our method attractive for fiber-based applications that require control over multiple inputs and/or wavelengths. 
Moreover, the possibility to achieve high dimension complex operations opens the way to the implementation of optical neural networks. 
Our system can provide an important building block for linear reconfigurable transformations, which can be further used in combination with fibers and lasers that exhibit strong gain and/or nonlinearity for deep learning applications.

\section*{Funding Information}

This research was supported by the Zuckerman STEM Leadership Program, the ISRAEL SCIENCE FOUNDATION (grant No. 1268/16), the Ministry of Science \& Technology, Israel and the French \textit{Agence Nationale pour la Recherche} (grant No. ANR-16-CE25-0008-01 MOLOTOF), and Laboratoire international associé Imaginano.

\section*{Acknowledgments}
We thank Daniel Golubchik and Yehonatan Segev for invaluable help. 

\section*{Disclosures}
The authors declare no conflicts of interest.

\printbibliography

\renewcommand{\thefigure}{S\arabic{figure}}
\setcounter{figure}{0}
\newpage
\section{Supplementary Material}

\subsection{Typical Time scales of the Optical System}

\subsubsection{Response Time}
  
\begin{figure*}[htb]
\begin{centering}
\includegraphics[width=0.7\linewidth]{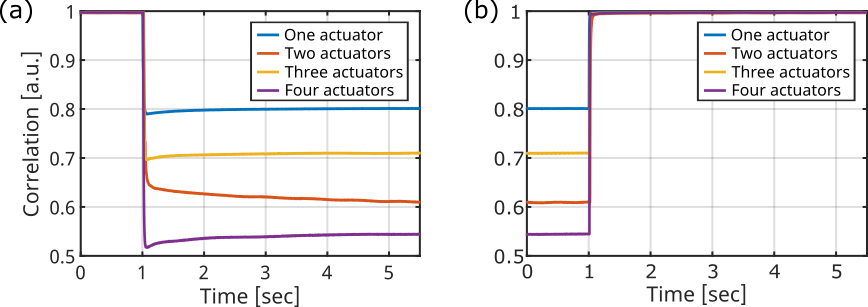}
\par\end{centering}
\caption{\label{fig:S1}\textbf{Response time of the experimental system.}
 The 2D correlation coefficient of each frame with (a) the first and (b) the last of the acquired frames, when a configuration of actuators (the voltage which is applied on these piezos) is changed. Blue lines show a change of configuration of a single actuator, red of two actuators, yellow of three and purple of four.} 
\label{fig:response}
\end{figure*}

To measure the typical response time of the system, we introduced abrupt changes to the voltages applied  to a subset of the piezoelectric actuators, and recorded  the speckle pattern obtained at the distal end of the fiber.
We then calculated the 2D Pearson correlation coefficient between each of the captured frames and the first frame. The measurements were repeated using different subsets of piezos. Examples of a few of these measurements, for subsets that include between one and four actuators, are shown in Fig. \ref{fig:response}(a).  The abrupt voltage change causes a fast change to the recorded speckle pattern, yielding a sharp decrease in the computed correlation coefficient. As expected, the bigger the subset of the piezos, the stronger the correlation drop. This sharp decrease is the result of the change in the actuators configuration (the bend they pose), and manifests in a change to the captured speckle pattern. 
Once the actuator's position stabilized, the correlation stabilized on a lower value. To ensure that the patterns with lower correlation with regard to the first frame are correlated with one another (thus ensuring that the plateau is not a result of the statistical properties of speckles), we also calculated the 2D correlation coefficient of each frame from the last acquired frame. These results are shown in Fig. \ref{fig:response}(b) for the same groups of actuators. The high correlation after the configuration change indeed verifies that the speckle pattern did not change further.
Based on such measurements we were able to estimate the response time of the system at $30 \hspace{2pt} ms$, which corresponds to modulation rates of 33 Hz.

\subsubsection{Decorrelation Time}
To estimate the stability of the system, we calculated the 2D correlation coefficient of the speckle pattern at the distal end of the fiber over time when the system is idle, i.e. no changes are performed to the states of the actuators. This loss of correlation is known to be attributed to the sensitivity of bare optical fibers to thermal fluctuations in the room and changes of pressure due to air flow. 
With the GRIN MMF, we found that the system remained highly correlated ($corr \geq 0.99$) for $\simeq10$ minutes. The correlation decreased slowly and linearly for 55 minutes, reaching $corr=0.976$. The correlation then decreased faster, reaching $corr=0.883$ after two hours. With the SI FMF, the system remained stable and highly correlated ($corr \geq 0.996$) over the course of 15 hours.

\subsection{Rayleigh Statistics}
\begin{figure}[htb]
\begin{centering}
\includegraphics[width=\columnwidth]{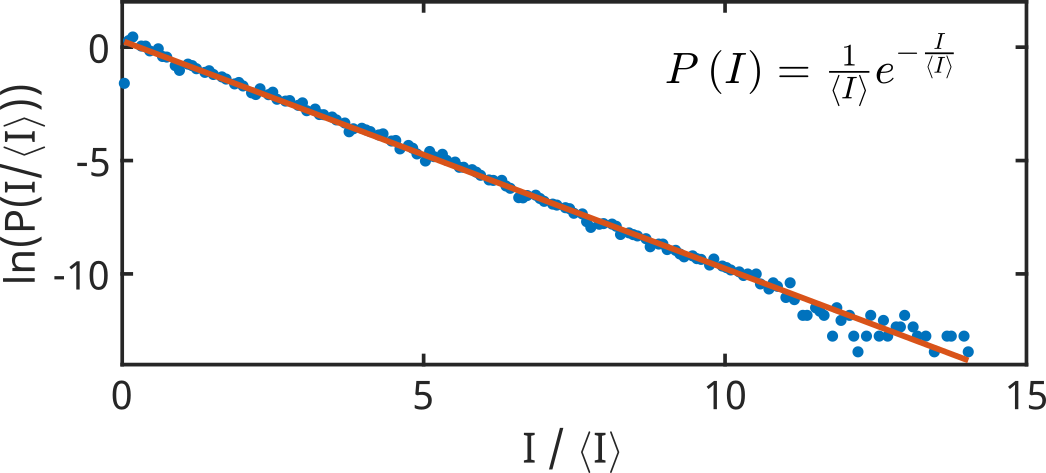}
\par\end{centering}
\caption{\textbf{Intensity distribution of the speckle patterns at the end of the fiber.}
 The natural logarithm of the probability distribution function (PDF) of the speckle patterns intensities, as a function of normalized pixel intensity (dots), and a linear fit (line). The data points correspond to experimental readouts, without background noise subtraction.}
\label{fig:rayleigh}
\end{figure}

The slopes of the linear scaling of the focusing enhancement factor $\eta$ as a function of the number of degrees of control rely on the intensity statistics of the generated speckle patterns. The theoretical values reported in the main text are derived for Rayleigh intensity statistics [1]. It is therefore important to compare the intensity statistics of the speckle patterns we obtain in our system with the predictions of Rayleigh statistics. Such a comparison is depicted in \ref{fig:rayleigh}, which shows excellent agreement with theory. 

\subsection{Optimization Technique}
As described in the main text, the results were obtained by finding solutions to optimization problems. These problems used a feedback loop- at each iteration, the speckle pattern at the distal end of the fiber was recorded using the CMOS camera.
This pattern was evaluated according to its similarity to a target pattern, and this score was given to the optimization algorithm as a cost, which it tried to minimize by changing the configuration of bends which are applied to the fiber segments. Lower costs were obtained for bend configurations which yielded patterns with high similarity to the target.

The optimization algorithm we chose to use is the Particle Swarm Optimization (PSO), which is a genetic algorithm. It randomly initializes a population of points (referred to as particles) in an \textit{M}-dimensional search space, representing the voltages which are assigned to the \textit{M} actuators. These positions are iteratively improved according to their local and global memory from previous iterations. Its stochastic nature helps avoiding local extrema in non-convex problems.
An open-source implementation of PSO [2] was modified to fit our experimental setup and simulation. We defined a single run as a single instance of the optimization process, i.e. achieving a single optimized target speckle pattern, such as the example shown in Fig. 3(b) in the main text. With the GRIN MMF, each such run constituted of 80 iterations, with the following hyper-parameters: population size of 120, inertia weight of $w=1$, inertia damping ratio of $w_{damp}=0.99$, personal learning coefficient of $c_1=1.5$, global learning coefficient of $c_2=2$. With the SI FMF, each run used 86-108 iterations, with a population of size 50. The values of the other hyper-parameters were not changed.

\subsection{Simulation}

Since our system is linear in the optical field, it is natural to describe the propagation of light in it with matrix formalism. We divided the fiber into multiple segments, calculated the transmission matrix (TM) of each segment and computed the total TM of the fiber by multiplying them. To represent our experimental system, we composed bent segments (which mimic the effect of actuators) and straight segments (for the propagation between actuators). A bent segment was approximated by a circular arc, with a defined curvature. To find the guided modes and propagation constants of different segments, we used a numerical module [3] which solves the scalar Helmholtz equation under the weakly guiding approximation [4]. We used 10 radii of curvatures, to simulate 10 different vertical positions of the actuators, which impose 10 different perturbations. These radii were linearly spaced between a maximal and a minimal value, which we estimated from the experimental system. 

Mode-mixing in short GRIN fibers mostly occurs within groups of degenerate modes. To mimic this phenomenon, we introduced unitary block matrices, whose block sizes were determined according to the mode degeneracy, as expressed in the propagation constants, allowing mixing between modes with the same propagation constants. It is note worthy that without introducing this feature, we were unable to achieve focusing. 

\begin{figure}[h]
\begin{centering}
\includegraphics[width=\linewidth]{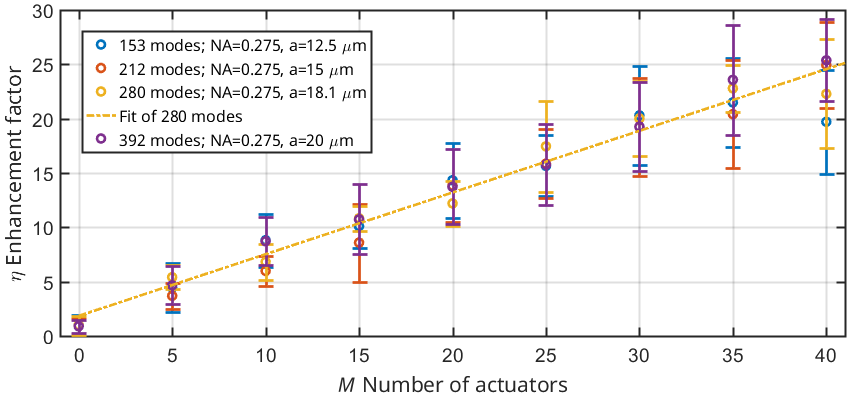}
\par\end{centering}
\caption{\textbf{Simulation of focusing in a multimode optical fiber.}
The average enhancement factor that was achieved (circles) and the standard error (bars), as a function of the number of actuators whose curvatures where modified as part of the optimization process in the simulation. Several results obtained for different fiber parameters are shown in different colors, along with a linear curve (gray dashed line).}
\label{fig:simulation}
\end{figure}

We used the same discrete set of possible curvatures for all of the actuators in all runs, and the same optimization mechanism as the experimental setup to achieve a focus. The optimization process mapped one of the possible curvatures for each of the bent fiber segments. In runs where not all of the actuators where used, the remaining were set as straight segments (with no curvature) to maintain the same propagation distance in all runs.
Fig. \ref{fig:simulation} shows the enhancement factor $\eta$ as a function of the number of actuators whose curvatures were optimized for simulated fibers. It is noticeable that the enhancement factor scales linearly with the number of simulated actuators, where the slope ranges between 0.57-0.64 for the displayed fiber parameters, at a wavelength of $\lambda=632.8 \hspace{2pt} nm$.

\end{document}